&pdflatex

\documentclass[prl, nofootinbib]{revtex4}

\global\arraycolsep=2pt
\usepackage{stmaryrd}
\usepackage{amsmath}
\usepackage{amssymb}
\usepackage{graphicx}
\usepackage{textcomp}
\usepackage{calrsfs}
\usepackage{yfonts}
\usepackage{bm}

\begin{document}

\title{Synthetic versus Dirac monopoles}

\author{Carl M. Bender$^1$, Maarten DeKieviet$^2$, and K. A. Milton$^3$}
\affiliation{
{}$^1$Department of Physics, Washington University, St. Louis, MO 63130, USA\\
{}$^2$Department of Physics and Astronomy, Heidelberg University, INF 226, 
69120 Heidelberg, Germany\\
{}$^3$Homer L. Dodge Department of Physics and Astronomy,
University of Oklahoma, Norman, OK 73019, USA}

\date\today
\begin{abstract}
In some recent experiments the distinction between synthetic magnetic monopoles
and Dirac monopoles has been blurred. A case in point is the work in a letter by
Ray {\it et al.}\ in which a beautiful experiment is reported but claims with
regard to Dirac monopoles are misleading.
\end{abstract}

\maketitle

A recent letter by Ray {\it et al.}\ \cite{R1} reports a nice experiment in
which from measured projected number densities the authors deduce what they call
a ``synthetic magnetic field.'' However, nothing in the experiment implies the
quantization of charge and thus the letter has no bearing on the Dirac monopole.
The wording in this letter is misleading; from a casual reading of the letter
one might come to the conclusion that the authors have actually found
experimental evidence for a Dirac monopole. In fact, the reported experiment
bears no serious relation to Dirac magnetic monopoles, whose necessary
attributes are described in detail in Ref.~\cite{R2}. There is no essential
gauge freedom, no charge quantization, and no invisible Dirac string. A Dirac
string is a pure gauge phenomenon and its spatial location can be changed by a
gauge transformation. Its location cannot be observed in any experiment. In
contrast, in the reported experiment there is a very visible vortex line, which
is a topological singularity in the sense that encircling it is associated with
a phase of $2\pi$. The interpretation of this phase is simply that it a Berry
phase, as discussed in detail nearly 30 years ago \cite{R3}. The term {\it
Berry phase} is not mentioned anywhere in the letter.

The physical observables discussed in the letter are the superfluid velocity
$\mathbf{v}_s$ and the vorticity $\bm{\Omega}=\bm{\nabla}\times\mathbf{v}_s$.
Dropping the primes used in the letter, the explicit forms for these quantities
for a vortex line starting at the origin and extending along the $z$ axis are
\begin{equation}
\mathbf{v}_s=\frac\hbar{Mr}\cot\frac\theta2 \hat{\bm\phi},\quad
\bm{\Omega}=-\frac\hbar{M}\left[\frac{\hat{\mathbf{r}}}{r^2}-\mathbf{f^*(r)}
\right].
\label{E1}
\end{equation}
Here the vortex line is described by
\begin{equation}
\mathbf{f^*(r)}=4\pi \delta(x)\delta(y)\theta(z)\hat{\mathbf{z}}.
\label{E2}
\end{equation}
Because $\bm{\nabla}\cdot\mathbf{f^*}=4\pi\delta({\mathbf{r}})$, it is indeed
true that $\bm{\nabla}\cdot\bm{\Omega}=0$. The content of the theoretical
analysis of the letter now seems to be the statement that
\begin{equation}
\bm{\Omega}=-\frac\hbar{M}(\mathbf{B^*-f^*}),\quad\bm{\nabla}\times\mathbf{v}_s
=-\frac\hbar{M}\bm{\nabla}\times \mathbf{A}^*,
\label{E3}
\end{equation}
where $\mathbf{B}^*=\mathbf{\hat r}/r^2$, or that 
\begin{equation}
\mathbf{B}^*=\bm{\nabla}\times\mathbf{A}^*+\mathbf{f^*}.
\label{E4}
\end{equation}
This indeed looks like the well-known equation relating the magnetic field
$\mathbf{B}$ of a unit point magnetic monopole at the origin to the gauge
potential $\mathbf{A}$ and the associated singular string function $\mathbf{f}$ 
\cite{R2}. However, this appearance is only formal. We can perform gauge
transformations on $\mathbf{A}^*$ but we cannot change $\mathbf{f}^*$ because it
refers to the location of the superfluid vortex line.

The transformations made in the supplementary information (19)--(21) of
Ref.~\cite{R1} are not permissible without changing the string function: {\it
The magnetic field of a magnetic monopole is not simply the curl of a vector
potential.} Since the ``string'' $\mathbf{f}^*$ here is a physical vortex line,
such a transformation is impossible. Because
\begin{equation}
\mathbf{A(r)}=\bm{\nabla}\lambda(\mathbf{r})-\frac1{4\pi}\int(d\mathbf{r'})
\mathbf{f(r-r')\times B(r')},\quad \lambda(\mathbf{r})=\frac1{4\pi}
\int(d\mathbf{r'})\mathbf{f(r-r')\cdot A(r')},
\label{E5}
\end{equation}
$\mathbf{A}$ is determined (up to a gradient) by specifying $\mathbf{f}$; thus,
with $\mathbf{f=f}^*$ in (\ref{E2}), the form for the vector potential given by
(\ref{E1}) does not possess any essential gauge freedom. That is, a
transformation $\mathbf{A}\to\mathbf{A}+\bm{\nabla}\lambda$ is permitted, but
the string cannot be rotated. In particular, the curl of the singular vector
potential $\tilde{\mathbf{A}}$ given in the letter's supplemental information
(19) cannot be equal to $\mathbf{B}^*$, because if it were, even in the
distributional sense, then we would obtain the result
\begin{equation}
\int_S d\mathbf{S}\cdot \mathbf{B}^*=\oint_{\partial S}d\mathbf{r}\cdot\mathbf{
\tilde{A}(r)}.
\label{E6}
\end{equation}
This equation cannot be true because if the surface $S$ were a boundary-free
spherical surface about the origin, the right side would be $4\pi$ while the
left side would vanish. Even if it were true that $\mathbf{B}^*=\bm{\nabla}
\times\tilde{\mathbf{A}}$, the physical vortex line $\mathbf{f}^*$ in (\ref{E4})
would still remain. Another reason why it is not true is that there is an
additional singularity, apparently unrecognized by the authors of
Ref.~\cite{R1}, which arises because the vector potential must be a
single-valued function. The string singularity is a necessary attribute of a
Dirac monopole, and no gauge transformation can ``annihilate'' it.

Moreover, a Dirac magnetic monopole is physically realizable in quantum
mechanics only because of the Dirac quantization condition between electric and
magnetic charge $e$ and $g$, $eg=n\hbar c$ (unrationalized units), where $n$ is 
an integer or an integer plus one-half. (It is precisely this condition that
makes the Dirac string invisible.) No analog of either electric or magnetic
charge appears in the analysis in the letter.

\acknowledgments
The work of CMB is supported in part by the US Department of Energy and that of
KAM by the Simons Foundation and the CNRS. KAM thanks Steve Fulling for
discussions.


\begin{thebibliography}{9}
\bibitem{R1} Ray, M. W., Ruokokoski, E., Kandel, S., M\"ott\"onen, M., and
Hall, D. S.  Observation of Dirac monopoles in a synthetic magnetic field.
Nature {\bf 505}, 657--660 (2014).
\bibitem{R2} Milton, K. A. Theoretical and experimental status of magnetic
monopoles. Rep.\ Prog.\ Phys.\ {\bf 69}, 1637--1711 (2006).
\bibitem{R3} M. Stone, Born-Oppenheimer approximation and the origin of
Wess-Zumino terms:  Some quantum-mechanical examples.
Phys.\ Rev.\ D {\bf 33}, 1191--1194 (1986).

\end{thebibliography}
\end{document}